\documentclass[10pt,twocolumn]{article}

\usepackage[utf8]{inputenc}
\usepackage{graphicx}
\usepackage{amssymb}
\usepackage{amsmath}
\usepackage[colorlinks=true,linkcolor=blue,citecolor=blue,urlcolor=blue]{hyperref}
\usepackage{xcolor}
\usepackage{ulem}

\usepackage[a4paper,margin=1.6cm]{geometry}
\usepackage{graphicx}
\usepackage{natbib}
\usepackage{abstract}

\graphicspath{Figures/}

\begin{document}

\title{Numerical simulations of oscillating and differentially rotating neutron stars}

\author{Santiago Jaraba$^{1,}$\thanks{santiago.jaraba-gomez@astro.unistra.fr}, J{\'e}r{\^o}me Novak$^{1,2}$ and Micaela Oertel$^{1,2}$ \\[2mm]
  {\small $^1$Observatoire astronomique de Strasbourg, CNRS, Université de Strasbourg, 11 rue de l'Université, 67000 Strasbourg, France}\\
{\small $^2$LUX, CNRS UMR 8262, Observatoire de Paris— PSL, Sorbonne Université Paris, 5 place Jules Janssen, 91190 Meudon, France}}

\twocolumn[
\maketitle
\begin{abstract}
    \noindent
    \textit{Context}. The remnants of binary neutron star mergers are expected to be massive, rapidly rotating stars whose oscillations produce gravitational waves in the kilohertz band. The degree of differential rotation and the rotation profiles strongly influence their structure, stability and oscillation spectrum, and must therefore be taken into account when modeling their dynamics.\\
    \textit{Aims}. We extend the pseudospectral code \texttt{ROXAS} (Relativistic Oscillations of non-aXisymmetric neutron stArS) to enable the dynamical evolution of oscillating, differentially rotating neutron stars. Using the updated code, we aim to study the star's oscillation frequencies.\\
    \textit{Methods}. We extend the previous formalism, based on primitive variables and the conformal flatness approximation, to differential rotation. Within this framework, we run a series of axisymmetric and non-axisymmetric simulations of perturbed, differentially rotating neutron stars with different rotation rates, and extract their oscillation frequencies. \\
    \textit{Results}. Axisymmetric modes, as well as those under the Cowling approximation, show excellent agreement with published results. We show that the secondary fundamental mode in the Cowling approximation is an artifact that does not appear in dynamical spacetimes. In addition, we provide, for the first time, frequency values for non-axisymmetric modes in differentially rotating configurations evolved in conformal flatness.\\
    \textit{Conclusions}. This extension broadens the range of physical scenarios that can be studied with \texttt{ROXAS}, and represents a step toward more realistic modeling of post-merger remnants and their gravitational-wave emission.
\end{abstract}
]

\begingroup
    \renewcommand\thefootnote{\fnsymbol{footnote}}
    \footnotetext[1]{\href{mailto:santiago.jaraba-gomez@astro.unistra.fr}{santiago.jaraba-gomez@astro.unistra.fr}}
\endgroup

\section{Introduction}

The direct detection of gravitational waves (GW), starting by the binary black hole merger GW150914~\citep{LIGOScientific:2016aoc}, has opened a new window onto the most extreme astrophysical phenomena. Subsequently, the first GW signal from a binary neutron star (NS) merger, GW170817~\citep{LIGOScientific:2017vwq}, was observed together with its electromagnetic counterpart~\citep{LIGOScientific:2017zic,LIGOScientific:2017ync}, providing unprecedented insight into the dynamics of this system~\citep{LIGOScientific:2018cki}. These detections were made possible by the current-generation GW detectors, LIGO~\citep{LIGOScientific:2014pky}, Virgo~\citep{VIRGO:2014yos} and KAGRA~\citep{KAGRA:2020tym}.

From the first numerical simulations of the merger of two NSs \citep{shibata-00b, shibata-05b} on, it is thought that in the majority of cases, the post-merger remnant of a binary NS is a hypermassive NS (HMNS)~\citep{Baumgarte:1999cq}, supported by its differential rotation for a few tenths of seconds. After that, the rotation would slow down and rigidify, making this object collapse to a black hole~\citep{Metzger:2019zeh}. During this period, the HMNS is expected to emit GWs in the kHz frequency band, in a very similar manner to excited differentially rotating isolated NSs~\citep{stergioulas-11}. Thus, the study of the properties of differentially rotating NSs \citep[see e.g. recent works by][]{weih-18, muhammed-24, szewczyk-25} is highly relevant to the knowledge of binary NS mergers~\citep{Bauswein:2015vxa}.

Note that current GW detectors lack sufficient sensitivity in the kHz band to detect the GWs from HMNS oscillations~\citep{KAGRA:2020}. However, next-generation detectors such as the Einstein Telescope~\citep{Punturo:2010zz}, Cosmic Explorer~\citep{Reitze:2019iox} and Neutron Star Extreme Matter Observatory~\citep{ackley-20} will have a significantly increased sensitivity in this frequency range, where future detections would be a probe of the NS internal structure and its equation of state (EoS). In order to maximize the scientific output from these signals by the time they arrive, it is thus necessary to have a solid understanding of how the EoS, rotation and other physical parameters shape the emitted GW.

To this end, the code \texttt{ROXAS} (Relativistic Oscillations of non-aXisymmetric neutron stArS)~\citep{Servignat:2024loh,servignat_2025_14849547} was recently developed to evolve isolated NSs. It is a full non-linear General-Relativistic hydrodynamics code, mostly aimed at obtaining oscillation modes from these objects \citep[see e.g.][for comparison with linear approaches]{baiotti-09}. It uses pseudospectral methods and a formulation based on primitive variables~\citep{Servignat:2022ovx} with the extended conformal flatness (xCFC) formulation developed in~\cite{Cordero-Carrion:2008grk}. The conformal flatness condition (CFC)~\citep{Isenberg:2007zg,Wilson:1996ty} has been proved to be a convenient approximation to full GR, allowing to simplify the formalism while keeping a high degree of accuracy~\citep{Iosif:2014dua}. These approaches reduce the computational cost with respect to other codes relying on conserved schemes~\citep{Banyuls:1997zz,Cipolletta:2019geh,Cipolletta:2020kgq,Dimmelmeier:2004me,Dimmelmeier:2005zk,Thierfelder:2011yi,Pakmor:2015ana,Lioutas:2022ghb,Kidder:2016hev,Rosswog:2020kwm,Yamamoto:2008js}, achieving a lightweight code that can be run on office computers and is thus ideally suited for parametric studies. In this paper, we present an update of \texttt{ROXAS} which allows to dynamically evolve perturbed, differentially rotating NSs. We describe the main changes to the formalism and code, and present a series of axisymmetric and non-axisymmetric simulations of differentially rotating stars known as the B sequence~\citep{Stergioulas:2003ep,Dimmelmeier:2005zk,Kruger:2009nw,Iosif:2020iho}, which use a polytropic EoS. We extract their oscillation modes and, when possible, compare their frequencies to those available in the literature for code validation, while the non-axisymmetric ones in CFC are reported for the first time.

The article is structured as follows. In section~\ref{sec:formalism}, we review the formalism used by \texttt{ROXAS} and stress the difference with differential rotation. In section~\ref{sec:simulations}, we present the changes implemented in the code and detail the parameters used in our simulations. The results are discussed in section~\ref{sec:results}, and finally conclusions with some final remarks are given in section~\ref{sec:conclusions}.

Unless otherwise stated, we use a geometrized unit system with $G=c=1$ and a 4-metric signature $(-,+,+,+)$. Greek indices denote the 4-dimensional spacetime coordinates, while Latin ones represent the 3-dimensional space. We also use Einstein's summation convention over repeated indices.

\section{Theoretical formalism}
\label{sec:formalism}

The formalism used here is based on the one in~\cite{Servignat:2024loh} and \cite{Servignat:2022ovx}, with some modifications to account for a differentially rotating star. The full details are provided in these references, and here we only summarize the main aspects and the key differences introduced in this paper.

\subsection{Review of formalism}

We consider a spacetime described by the 3+1 formalism, so that the metric takes the form
\begin{equation}
    g_{\mu\nu}dx^\mu dx^\nu=-N^2dt^2+\gamma_{ij}(dx^i+\beta^i dt)(dx^j+\beta^j dt),
\end{equation}
where $N$ is the lapse, $\beta^i$ the shift and $\gamma_{ij}$ the induced spatial metric. We assume CFC, so that
\begin{equation}
    \gamma_{ij}=\Psi^4 f_{ij},
\end{equation}
where $\Psi$ is called conformal factor and $f_{ij}$ is a flat metric. We denote the covariant derivative associated to $\gamma_{ij}$ as $D_i$ and, as described by \cite{bonazzola-04}, we work in maximal slicing and the Dirac gauge, which is automatically satisfied in CFC. The flat metric $f_{ij}$ shall be expressed in spherical coordinates $\left(r, \theta, \varphi \right)$, as in \cite{Servignat:2024loh}.

On the hydrodynamics side, we consider a simulated NS to be a perfect fluid, whose energy-momentum tensor is given by
\begin{equation}
    T^{\mu\nu}=(e+p)u^\mu u^\nu+pg^{\mu\nu},
\end{equation}
where $u^\mu$ is the timelike unitary 4-velocity, $e$ is the energy density in the fluid frame and $p$ is the pressure. The fluid is composed of baryons with number density $n_B$, at zero temperature and in $\beta$ equilibrium. The fluid thermodynamics can then be described by a barotropic EoS, $p=p(n_B)$, which we will assume to be polytropic throughout this work, that is,
\begin{equation}
    p=\kappa n_B^\gamma,
\end{equation}
where $\gamma=2$ and $\kappa=0.02689$ in \texttt{LORENE} units\footnote{This corresponds to the standard polytropic EoS in e.g.~\cite{Stergioulas:2003ep,Dimmelmeier:2005zk}, where $p=\kappa_\rho\rho^\gamma$, with $\rho = m_B n_B$, $\gamma=2$ and $\kappa_\rho=100$ in the geometrized unit system supplemented with $M_\odot=1$.}, defined by $c=r_0=m_0=1$, with $r_0=10$ km and $m_0\approx 1.66\times 10^{29}$ kg. We also define the log-enthalpy $H$,
\begin{equation}
    H=\ln\left(\frac{e+p}{m_Bn_B}\right),
\end{equation}
where $m_B\approx1.66\times 10^{-27}$ kg is the nucleon mass. Additionally, we consider the Lorentz factor with respect to the Eulerian observer $\Gamma$, the Eulerian velocity of the fluid $U^i$ and its coordinate velocity $v^i$, which satisfy the following equations:
\begin{equation}
    \label{eq:Gamma}
    \Gamma = (1-U_jU^j)^{-1/2},
\end{equation}
\begin{equation}
    u^\alpha=\frac{\Gamma}{N}(1,v^i),\quad U^i=\frac{1}{N}(v^i+\beta^i).
\end{equation}

Within this framework, the basic evolution equations are~\citep{Servignat:2022ovx,Servignat:2024loh}
\begin{align}
    \label{eq:H evol}
    \partial_t H = -v^iD_iH&-c_s^2\frac{\Gamma^2N}{\Gamma^2-c_s^2(\Gamma^2-1)}\\
    \nonumber
    &\times\left[K_{ij}U^iU^j+D_iU^i-\frac{U^i}{\Gamma^2}D_iH\right],
\end{align}
\begin{align}
    \nonumber
    \partial_t U_i =& -v^jD_jU_i-U_jD_i v^j+NU_jD_iU^j-\frac{N}{\Gamma^2}D_i(H+\ln N)\\
    \label{eq:Ui evol}
    &+U_iU^jD_jN+\frac{c_s^2NU_i}{\Gamma^2-c_s^2(\Gamma^2-1)}D_jU^j\\
    \nonumber
    &+U_i\frac{\Gamma^2(c_s^2-1)}{\Gamma^2-c_s^2(\Gamma^2-1)}NU^lU^jK_{lj}\\
    \nonumber
    &+\frac{(1-c_s^2)N}{\Gamma^2-c_s^2(\Gamma^2-1)}U_iU^jD_jH.
\end{align}

Similarly to~\cite{Servignat:2024loh}, we decompose each field into the equilibrium value, denoted with a subscript ``eq'', plus the deviation of the total quantity from this equilibrium, denoted with a bar,
\begin{align}
    U_i&=U_{i,\rm eq}+\bar{U}_i,\quad H=H_{\rm eq}+\bar{H},\\
    \nonumber
    \beta^i&=\beta^i_{\rm eq}+\bar{\beta}^i,\quad N=N_{\rm eq}+\bar{N}.
\end{align}
Note that we do not assume that the latter quantities need to be small.

The next step is to substitute these expressions into Eqs.~\eqref{eq:H evol}, \eqref{eq:Ui evol}, and use the identities satisfied at equilibrium to simplify the resulting equations. This is what we call ``well-balanced formulation'', and where the main differences between rigid and differential rotation will arise. However, this is not strictly speaking a well-balanced method as defined e.g. by \cite{dumbser-23}, but it is inspired from it. 

\subsection{Differential rotation}

An axisymmetric star at equilibrium with an angular velocity $\Omega(r,\theta)$, so that $v_{\rm eq}^i=\Omega(r,\theta)r\sin\theta\,\delta^i_\varphi$, satisfies the equation~\citep{Bonazzola:1993zz,Iosif:2020iho}
\begin{equation}
    \label{eq:equilibrium_identity}
    H_{\rm eq}+\ln N_{\rm eq}-\ln\Gamma_{\rm eq}+\int_{\Omega_p}^\Omega F(\Omega')d\Omega'={\rm const.},
\end{equation}
where
\begin{equation}
    F=u^tu_\varphi=\frac{\Gamma_{\rm eq}^2}{N_{\rm eq}} U_{\varphi,\rm eq}\,r\sin\theta
\end{equation}
and $\Omega_p$ is the angular velocity at the star's pole.

The introduction of differential rotation thus involves defining a rotation law relating these two variables, $F(\Omega)$. Historically, due to its simplicity, the most standard choice has been the Komatsu-Eriguchi-Hachisu (KEH) profile~\citep{Komatsu:1989ikr},
\begin{equation}
    \label{eq:KEH_F_Om}
    F(\Omega)=A^2(\Omega_c-\Omega),
\end{equation}
where $\Omega_c$ is the central angular velocity and $A$ is a constant which determines the length scale over which the angular velocity $\Omega$ changes. This can easily be seen in the Newtonian limit, where $u^t\approx 1$, $g_{\varphi\varphi}\approx(r\sin\theta)^2$ and $u^\varphi\approx\Omega$, getting the simple rotation law
\begin{equation}
    \Omega(r,\theta)_{\rm Newt}=\frac{\Omega_c}{1+(r\sin\theta/A)^2}.
\end{equation}
In particular, $A\rightarrow\infty$ corresponds to the rigidly rotating limit. The KEH profile produces a maximum frequency at the center of the star, which monotonically decreases with the radius, as can later be seen in Fig.~\ref{fig:rot_profiles}.

More recently, other differential rotation laws have been proposed which more realistically describe the situation in the post-merger remnant, such as those in~\cite{Uryu:2017obi}. However, these profiles introduce additional computational challenges, as discussed in Section~\ref{sec:simulations}. In addition, for code validation purposes, it is useful to consider rotation profiles which have been more broadly studied in previous references. For these reasons, in the present paper, our numerical simulations will only deal with KEH profiles, leaving more realistic and complex rotation laws for future work. In any case, the formalism described in this section remains general, with no particular assumption on the shape of $F(\Omega)$.

\subsection{The well-balanced formulation with differential rotation}

One of the main changes to obtain the hydrodynamics equations with differential rotation in the well-balanced formulation is the integral term appearing in Eq.~\eqref{eq:equilibrium_identity}. The other relevant identity which impacts the dynamical equations is
\begin{equation}
    \label{eq:vdu_eq}
    v_{\rm eq}^jD_jU_{i,\rm eq}+U_{j,\rm eq}D_iv_{\rm eq}^j=U_{\varphi,\rm eq}\,r\sin\theta D_i\Omega,
\end{equation}
with the right-hand side vanishing for rigid rotation. Both these equations produce additional terms in Eq.~\eqref{eq:Ui evol} with respect to the rigidly rotating case, leading to the final evolution equation for $U_i$:
\begin{align}
    \nonumber
    \partial_t\bar{U}_i =& -\biggl[v_{\rm eq}^jD_j\bar{U}_i+\bar{v}^jD_jU_{i,\rm eq}+\bar{v}^jD_j\bar{U}_i\\
    \nonumber
    &+U_{j,\rm eq}D_i\bar{v}^j+\bar{U}_jD_iv_{\rm eq}^j+\bar{U}_jD_i\bar{v}^j\\
    \nonumber
    &-N\left(U_{j,\rm eq}D_i\bar{U}^j+\bar{U}_jD_iU^j_{\rm eq}+\bar{U}_jD_i\bar{U}^j\right)\biggr]\\
    \nonumber
    &+N\bar{U}_j(2U^j_{\rm eq}+\bar{U}^j)D_i(H_{\rm eq}+\ln N_{\rm eq})\\
    \nonumber
    &-\frac{N}{\Gamma^2}D_i\left(\bar{H}+\ln \left(1+\frac{\bar{N}}{N_{\rm eq}}\right)\right)+U_iU^jD_jN\\
    \nonumber
    &+\frac{c_s^2NU_i}{\Gamma^2-c_s^2(\Gamma^2-1)}(6U_{\rm eq}^\varphi\bar{D}_\varphi\ln\Psi+D_j\bar{U}^j)\\
    \nonumber
    &+U_i\frac{\Gamma^2(c_s^2-1)}{\Gamma^2-c_s^2(\Gamma^2-1)}NU^lU^jK_{lj}\\
    \nonumber
    &+\frac{(1-c_s^2)N}{\Gamma^2-c_s^2(\Gamma^2-1)}U_iU^jD_jH\\
    \label{eq:Ui_evol_WB}
    &+\frac{\bar{N}}{N_{\rm eq}}U_{\varphi,\rm eq}\,r\sin\theta\,D_i\Omega,
\end{align}
where the last term is the combination of the two described contributions from differential rotation. Two additional terms appear which were not present in~\cite{Servignat:2024loh}, corresponding to higher-order contributions that do not contribute to the evolution in a relevant way in most cases. However, it is safer to account for them, especially for high velocities or if large deviations from equilibrium are involved.

Regarding the equation for the enthalpy, Eq.~\eqref{eq:H evol} can be slightly simplified by using the same identity used for rigid rotation, $v^i_{\rm eq}D_iH_{\rm eq}=0$, leading to the same evolution equation for $H$:
\begin{align}
    &\partial_t\bar{H}=-\bar{v}^iD_iH_{\rm eq}-v_{\rm eq}^iD_i\bar{H}-\bar{v}^iD_i\bar{H}\\
    \nonumber
    &-c_s^2\frac{\Gamma^2N}{\Gamma^2-c_s^2(\Gamma^2-1)}\left[K_{ij}U^iU^j-K+D_iU^i-\frac{U^i}{\Gamma^2}D_iH\right].
\end{align}

\section{Numerical simulations}
\label{sec:simulations}

The numerical simulations performed in this work have been done with a modified version of the code \texttt{ROXAS}~\citep{Servignat:2024loh}~\footnote{The version with rigid rotation was made publicly available in~\cite{servignat_2025_14849547} in February 2025.}. The changes made to the code -- mainly motivated by the introduction of differential rotation -- are summarized below.

\subsection{Updates to the code}

\texttt{ROXAS} is a pseudospectral code for the dynamical evolution of perturbed, isolated rotating NSs. It makes use of the \texttt{LORENE} (Langage Objet pour la RElativité NumériquE)~\citep{LORENE} infrastructure and, in particular, it relies on it to produce the equilibrium star that will be perturbed and evolved. Therefore, differential rotation needs to be implemented in both \texttt{LORENE} and \texttt{ROXAS}.

Although tools for computing the equilibrium configuration of a differentially rotating star have previously been implemented in \texttt{LORENE}~\citep{Saijo:2006um}, they were not under the CFC assumption and needed to be updated.

Despite the fact that we focus on KEH rotation profiles throughout this text, the goal is to be able to use more complex profiles in the future. Profiles such as those in~\cite{Uryu:2017obi} need to be specified through a non-injective function $\Omega(F)$, which cannot be inverted as $F(\Omega)$. While KEH profiles are generally defined through the $F(\Omega)$ in Eq.~\eqref{eq:KEH_F_Om}, this relation can be inverted and expressed also as an $\Omega(F)$,
\begin{equation}
    \label{eq:KEH_Om_F}
    \Omega(F)=\Omega_c-\frac{F}{A^2}.
\end{equation}

With this change, the integral term in the equilibrium equation~\eqref{eq:equilibrium_identity} needs to be computed as
\begin{equation}
    \int_{F_p}^F F'\frac{d\Omega}{dF}(F')\,dF',
\end{equation}
but otherwise the formalism remains the same. As using $\Omega(F)$ is therefore a more generic framework, we chose this way to define the rotation profile for our implementation, in contrast to the original implementation of differential rotation in \texttt{LORENE}.

Note that this step alone is insufficient to deal with the profiles in~\cite{Uryu:2017obi}, since other challenges remain. For instance, these profiles involve a higher number of free parameters whose values are usually determined through other variables with more clear physical meaning, as can be seen in~\cite{Iosif:2020iho}. This implies that these parameters change during the iterative process to find the star's equilibrium configuration, possibly leading to failure of the algorithm or to non-physical configurations if not treated carefully, especially in the most rapidly rotating cases. Another issue is that these profiles show a local maximum in the angular velocity at a non-zero radius, contrary to KEH profiles (see Fig.~\ref{fig:rot_profiles}), which requires more resolution around this point in the dynamical evolution and usually leads to less stable simulations.

Once a NS equilibrium configuration is found, the resulting fields are stored for later usage with \texttt{ROXAS}, including $\Omega$ (for usage e.g. in Eq.~\eqref{eq:Ui_evol_WB}) instead of $F$, which can be recovered from other variables.

\texttt{ROXAS} was also modified to read equilibrium configurations of differentially rotating stars, as well as to account for the additional terms described in section~\ref{sec:formalism} for a consistent dynamical evolution. Before evolving the star, a perturbation is added either to the Eulerian velocity $U_i$ or to the log-enthalpy $H$. In the latter case, the perturbation can be chosen to be
\begin{equation}
    \label{eq:delta_H}
    \delta H=\varepsilon\left(x^2\pm y^2-\frac{2}{3}r^2\right)\left(1-\frac{r^2}{R_S(\theta)^2}\right),
\end{equation}
where $R_S(\theta)$ is the radius of the equilibrium star in the direction of $\theta$. In this equation, the $-$ sign corresponds to a non-axisymmetric perturbation mixing $l=m=0$ and $l=m=2$ modes, as defined and used in~\cite{Servignat:2024loh}, while the $+$ sign yields axial symmetry, useful to save simulation time if only an axisymmetric study is required.

\begin{figure}[h]
    \centering
    \includegraphics[width=\linewidth]{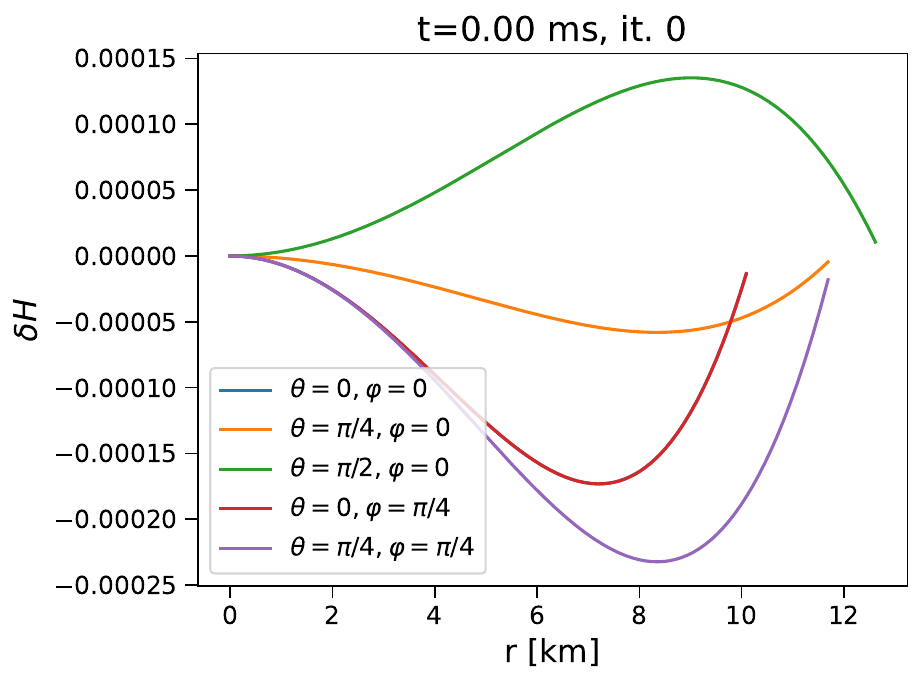}
    \caption{Radial profile in different directions of the initial enthalpy perturbation of the non-axisymmetric B4 simulations.}
    \label{fig:ent_prof}
\end{figure}

\begin{figure}[h]
    \centering
    \includegraphics[width=\linewidth]{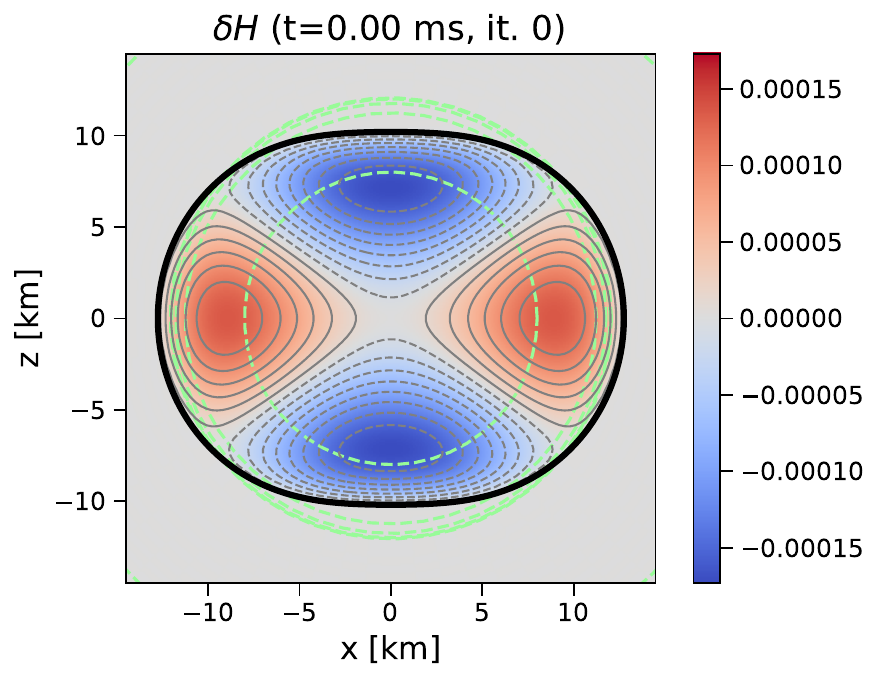}
    \caption{Slice in the $xz$ plane of the initial enthalpy perturbation of the non-axisymmetric B4 simulations. The green dashed lines correspond to the boundaries between the different domains of the metric grid, as described in Subsection~\ref{ssec:sim_params}.}
    \label{fig:ent_slice}
\end{figure}

Finally, \texttt{ROXAS} is now able to export 1-dimensional profiles and 2-dimensional slices of the hydrodynamic and metric grids during the simulation, where the output rate can be specified independently. They can easily be plotted afterwards, allowing to have a better understanding of how the simulation internally behaves, and to spot potential issues. Examples with the initial profile and slice of the enthalpy perturbation of a non-axisymmetric simulation are shown in Figs.~\ref{fig:ent_prof} and~\ref{fig:ent_slice}, respectively.

\subsection{Simulation parameters}
\label{ssec:sim_params}

In order to test the new implemented changes, we check the consistency with other numerical simulations of oscillations in differentially rotating stars in the literature. In particular, we use the B sequence described in~\cite{Dimmelmeier:2005zk} and compare the resulting frequencies with those provided in this reference.

For the B sequence, the central value for the log-enthalpy is the same as in the BU sequence used in~\cite{Servignat:2024loh}, which in \texttt{LORENE} units is $H_c=0.2280$. For the rotation profile in Eq.~\eqref{eq:KEH_F_Om}, we set $A=r_e$, with $r_e$ the star radius at the equator. The B sequence is defined by the star's flattening (ratio between the star radius at the pole, $r_p$, and $r_e$). However, \texttt{LORENE} does not take this as an input parameter, and requires the central angular velocity $\Omega_c$. Therefore, we take these values from~\cite{Dimmelmeier:2005zk} and check that the final configurations found by the algorithm yield similar values of $r_p/r_e$, masses and ratios between kinetic and potential energies, $T/W$. A summary of these parameters and their relative errors with respect to those in~\cite{Dimmelmeier:2005zk} is provided in Table~\ref{tab:initial_pars}. We can see that the agreement is excellent, even for the fastest rotating stars. The largest errors correspond to $T/W$ for B1 and B2, due to the smallness of these quantities compared to the precision of the reference values.

\begin{table*}[h]
    \centering
    \begin{tabular}{|c||c|c|c|c|c||c|c|c|c|}
        \hline
        Model & $f_c$ (Hz) & $f_e$ (Hz) & $M~(M_\odot)$ & $r_p/r_e$ & $T/W$ & $\delta f_e$ (\%) & $\delta M$ (\%) & $\delta(r_p/r_e)$ (\%) & $\delta(T/W)$ (\%)\\\hline
        B0 & 0 & 0 & 1.400 & 1.000 & 0.000 & 0.00 & 0.01 & 0.00 & 0.00 \\
        B1 & 582 & 215 & 1.437 & 0.950 & 0.012 & 0.00 & 0.03 & 0.00 & 3.85 \\
        B2 & 832 & 305 & 1.478 & 0.900 & 0.026 & 0.06 & 0.02 & 0.00 & 1.02 \\
        B3 & 1030 & 375 & 1.525 & 0.850 & 0.040 & 0.02 & 0.03 & 0.01 & 0.58 \\
        B4 & 1205 & 434 & 1.578 & 0.800 & 0.055 & 0.07 & 0.02 & 0.01 & 0.67 \\
        B5 & 1366 & 486 & 1.640 & 0.750 & 0.070 & 0.09 & 0.02 & 0.01 & 0.81 \\
        B6 & 1521 & 534 & 1.713 & 0.700 & 0.087 & 0.16 & 0.01 & 0.03 & 0.24 \\
        B7 & 1675 & 579 & 1.798 & 0.650 & 0.105 & 0.17 & 0.00 & 0.03 & 0.03 \\
        B8 & 1836 & 622 & 1.900 & 0.600 & 0.124 & 0.23 & 0.03 & 0.07 & 0.01 \\
        B9 & 2014 & 665 & 2.022 & 0.549 & 0.144 & 0.30 & 0.08 & 0.09 & 0.05 \\\hline
    \end{tabular}
    \caption{Central ($f_c$) and equatorial ($f_e$) frequencies, gravitational mass ($M$), flattening ($r_p/r_e$) and ratio between kinetic and potential energies ($T/W$) of the computed NS equilibrium configurations, together with their relative errors with respect to the reference values from~\cite{Dimmelmeier:2005zk}. Note that there is no error for $f_c$, as we take it as an input parameter.}
    \label{tab:initial_pars}
\end{table*}

With respect to the dynamical evolution with \texttt{ROXAS}, the simulation properties are very similar to the ones in~\cite{Servignat:2024loh}. Two types of grids are used, with the following parameters:
\begin{itemize}
    \item The hydro grid consists of two domains: the spherical nucleus, with a maximum radius of $0.7r_p$, and a shell that covers the rest of the star. In both domains, we set $N_r=N_\theta=17$, with $N_i$, $i=r,\theta,\varphi$, the number of grid points for each coordinate. In addition, $N_\varphi$ is set to 8 for the non-axisymmetric simulations and to 1 for the axisymmetric ones. In~\cite{Servignat:2024loh}, $N_\theta$ is increased for the most demanding (i.e. more rapidly rotating) simulations, but we found them to be more stable by keeping the same value.

    \item The metric grid consists of several spherical domains: a nucleus that extends up to 8 km (7 km for B9, which has $r_p\approx 7.5$ km), a first shell from this radius up to $r_p$, four more shells up to $r_e$, and then one more up to 20 km followed by the compactified external domain, in which a change of variable $u=1/r$ is done to take into account boundary conditions at $r \to \infty$ \citep{bonazzola-98}. The four shells between $r_p$ and $r_e$ are chosen so that their edges match $R_S(\theta)$ for $\theta$ uniformly spaced between 0 and $\pi/2$. 
\end{itemize}

\section{Results}
\label{sec:results}

\begin{figure}[h]
    \centering
    \includegraphics[width=\linewidth]{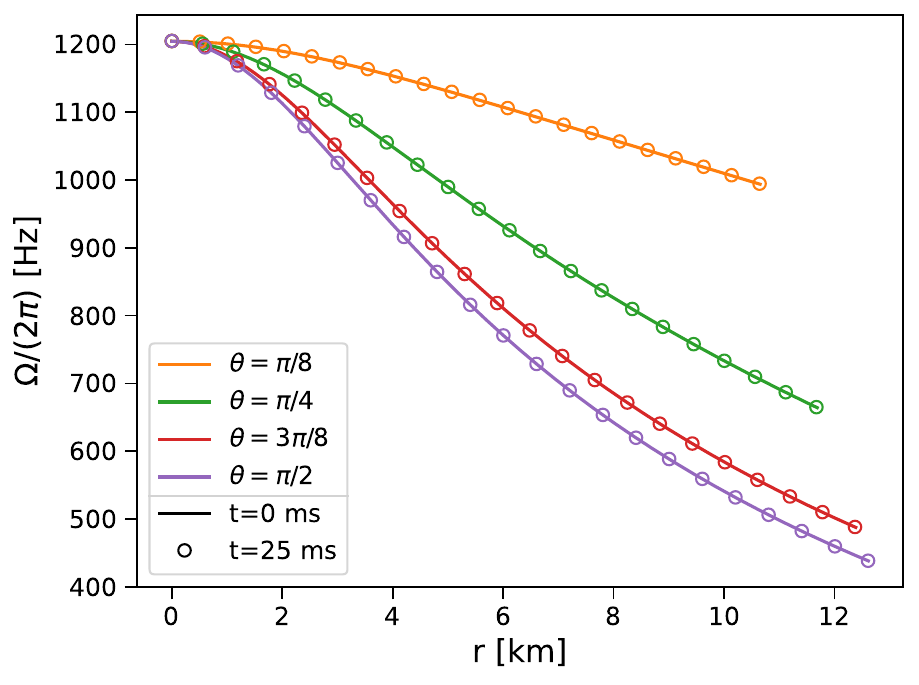}
    \caption{Angular velocity profiles in different directions for an unperturbed B4 simulation. The solid lines correspond to the initial profiles, while the superimposed points are plotted after 25 ms of dynamical evolution in CFC.}
    \label{fig:rot_profiles}
\end{figure}

A first test of the code was done to check its ability to keep stable equilibrium initial data, with the evolution of unperturbed differentially rotating NS. Some angular velocity profiles corresponding to an unperturbed B4 are thus shown in Fig.~\ref{fig:rot_profiles}, both initially and after a dynamical evolution in CFC of 25 ms. The fact that these profiles remain constant after this time period (many roation periods) shows that the formalism, the dynamical code and the equilibrium star solvers work together consistently and in a stable way.

Then, for each of the differentially rotating NS in equilibrium reported in Table~\ref{tab:initial_pars}, we perform four simulations: two in the Cowling approximation and another two with dynamical spacetime under CFC. For each pair, one of the simulations is evolved in axisymmetry after adding an axisymmetric perturbation of the form~\eqref{eq:delta_H} ($+$ sign) to the equilibrium star, while the other one is evolved in 3D\footnote{We impose symmetries about the equatorial plane and by rotation of $\pi$ around the vertical z axis. These symmetries include the dominant quadrupolar $\ell = |m| = 2$ mode (see \cite{Servignat:2024loh})} after adding a non-axisymmetric perturbation of the form~\eqref{eq:delta_H} ($-$ sign). In all cases, the amplitude is chosen to be $\epsilon=10^{-3}$ in code units.

The gravitational waves are extracted following the procedure in~\cite{Servignat:2024loh}, based on Einstein's quadrupole formula, where an example waveform is later provided in Fig.~\ref{fig:B4_GW}.

\subsection{Cowling approximation}

The Cowling approximation assumes that the metric stays static and equal to the values at equilibrium. This approximation has been widely used for the simulation of relativistic stars~\citep[see e.g.][]{mcdermott-83, villain-05, sotani-23, montefusco-25} due to the reduction of the computational cost of simulations. While this is not particularly relevant for \texttt{ROXAS}, running some simulations in the Cowling approximation allows us to perform more extensive code validation against existing literature. In particular, axisymmetric modes can be compared to the ones in~\cite{Stergioulas:2003ep}, which uses an axisymmetric code for dynamical evolution in General Relativity under the Cowling approximation. On the other hand, non-axisymmetric modes can be compared to~\cite{Kruger:2009nw}, where these frequencies are obtained with a perturbative approach, once again limited to the Cowling approximation.

\begin{figure}[h]
    \centering
    \includegraphics[width=\linewidth]{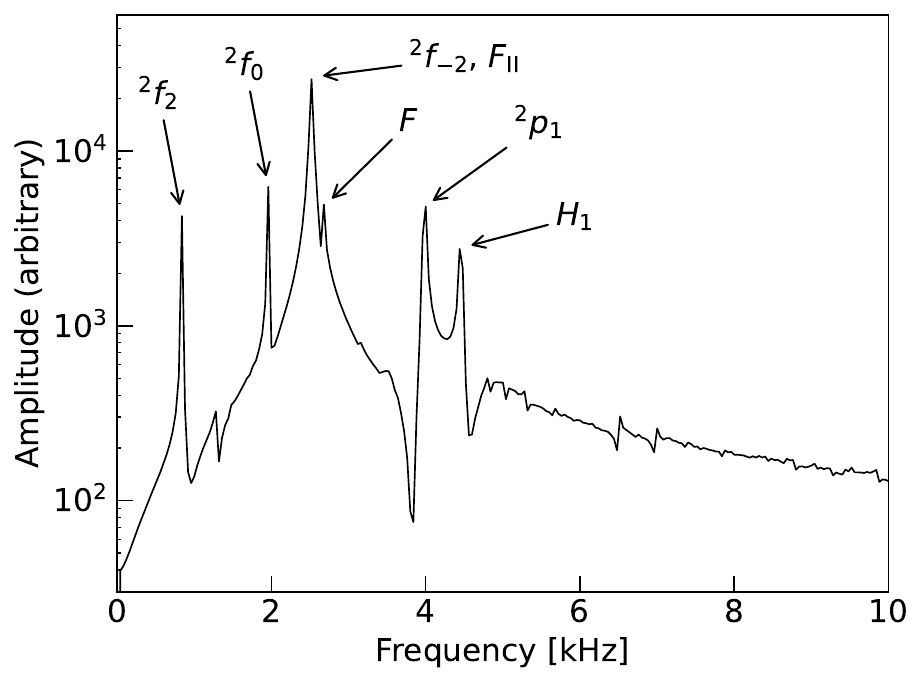}
    \caption{GW spectrum of the non-axisymmetric B4 simulation in the Cowling approximation.}
    \label{fig:B4_Cowl_spectrum}
\end{figure}

When possible, these simulations are run for 25 ms; however, for the most demanding configurations, the non-axisymmetric simulations are only stable up to shorter times, which are reported in Table~\ref{tab:freqs_cowling}.
An example frequency spectrum for the non-axisymmetric B4 simulation is provided in Fig.~\ref{fig:B4_Cowl_spectrum}, where we can distinguish the different oscillation modes present in the simulation. In particular, a secondary fundamental mode $F_{\rm II}$ appears around 2.6 kHz, as was reported for the first time in~\cite{Stergioulas:2003ep}. As $F_{\rm II}$ and $^2f_{-2}$ have similar values, they generate a single joint peak in Fig.~\ref{fig:B4_Cowl_spectrum}. However, this secondary fundamental mode can be identified independently in the absence of an $^2f_{-2}$, such as in axisymmetric simulations or in spectra corresponding to $l=m=0$ modes, as in Fig.~\ref{fig:B4_00_spectra} presented in the next subsection.

The frequencies for the full B sequence are presented in Table~\ref{tab:freqs_cowling}, where the precise values are extracted from the Fourier transform of the decomposition of the star's radius $R_S$ into spherical harmonics. These time series include less oscillation modes per multipole order, leading to more isolated and cleaner peaks and, therefore, to a more precise frequency determination.

The axisymmetric frequencies we extract are the fundamental modes $F$ and $F_{\rm II}$, their first overtone $H_1$, the $l=2,m=0$ mode ${}^2f_0$ and its first overtone $^2p_1$. The values shown in Table~\ref{tab:freqs_cowling} correspond to those obtained from the axisymmetric simulations, given their better stability and less number of modes, which generate cleaner peaks. However, as these frequencies are also present in the non-axisymmetric simulations, we extract them from these simulations as well, obtaining the same values up to maximum relative errors of 0.9\%. On the other hand, the extracted non-axisymmetric frequencies are the $l=|m|=2$ modes, denoted by ${}^2f_2$ and ${}^2f_{-2}$, and appear only in the non-axisymmetric simulations.

\begin{table*}[h]
    \centering
    \centerline{
    \begin{tabular}{|c||c|c|c|c|c|c|c||c|c|c|c|c|c|c||c|}
        \hline
         & $F_{\rm II}$ & $F$ & $H_1$ & ${}^2f_0$ & $^2p_1$ & ${}^2f_2$ & ${}^2f_{-2}$ & $\delta F_{\rm II}$ & $\delta F$ & $\delta H_1$ & $\delta\,{}^2f_0$ & $\delta\,{}^2p_1$ & $\delta\,{}^2f_2$ & $\delta\,{}^2f_{-2}$ & $t_{\rm 3D}$ \\\hline
        B0 & 2.680 & 2.680 & 4.557 & 1.881 & 4.116 & 1.881 & 1.881 & 0.95 & 0.95 & 0.23 & 1.87 & 0.38 & 0.16 & 0.16 & 25.0 \\
        B1 & 2.663 & 2.663 & 4.487 & 1.917 & 4.109 & 1.439 & 2.242 & 1.35 & 0.17 & 0.93 & 1.98 & 0.18 & 0.74 & 0.13 & 25.0 \\
        B2 & 2.642 & 2.642 & 4.471 & 1.922 & 4.081 & 1.201 & 2.364 & 3.16 & 0.18 & 1.12 & 1.14 & 0.22 & 0.22 & 0.03 & 25.0 \\
        B3 & 2.555 & 2.649 & 4.443 & 1.953 & 4.041 & 1.011 & 2.450 & 1.18 & 0.65 & 0.85 & 2.10 & 0.11 & 0.40 & 0.03 & 25.0 \\
        B4 & 2.552 & 2.678 & 4.448 & 1.960 & 3.971 & 0.840 & 2.522 & 1.85 & 1.74 & 1.01 & 1.86 & 0.30 & 0.07 & 0.28 & 25.0 \\
        B5 & 2.519 & 2.680 & 4.480 & 1.960 & 3.909 & 0.679 & 2.563 & 1.29 & 1.84 & 1.32 & 1.62 & 0.04 & 0.95 & 0.15 & 25.0 \\
        B6 & 2.481 & 2.684 & 4.488 & 1.959 & 3.804 & 0.517 & 2.623 & 0.88 & 1.93 & 1.18 & 1.77 & 0.63 & 2.42 & 0.46 & 21.1 \\
        B7 & 2.451 & 2.701 & 4.519 & 1.933 & 3.718 & 0.331 & 2.651 & 1.15 & 2.49 & 1.61 & 1.24 & 0.07 & 2.12 & 0.10 & 21.2 \\
        B8 & 2.437 & 2.717 & 4.507 & 1.920 & 3.629 & 0.174 & 2.694 & 1.79 & 2.70 & 1.41 & 1.61 & 0.07 & 7.12 & 0.43 & 17.2 \\
        B9 & 2.398 & 2.720 & 4.442 & 1.884 & 3.543 & -- & 2.750 & 1.62 & 2.54 & 0.65 & 0.92 & 0.67 & -- & 0.88 & 13.2 \\\hline
    \end{tabular}}
    \caption{Frequencies (kHz) extracted from the B sequence simulations in the Cowling approximation and their relative errors (\%) with respect to the results in~\cite{Stergioulas:2003ep} (axisymmetric modes) and~\cite{Kruger:2009nw} (non-axisymmetric ones). The last column indicates the total simulation time for the non-axisymmetric simulations in milliseconds, while the axisymmetric ones ran for 25 ms.}
    \label{tab:freqs_cowling}
\end{table*}

The relative errors of our results with respect to~\cite{Stergioulas:2003ep} and~\cite{Kruger:2009nw} are provided in Table~\ref{tab:freqs_cowling}. Most of them are similar or below 2\%. The larger values of $\delta\,{}^2f_2$ from B6 on can be explained by the smallness of these frequencies compared to our frequency resolution. Similarly, the 3.16\% error in $F_{\rm II}$ for B2 is due to the fact that the peaks for $F$ and $F_{\rm II}$ are completely blended in our B2 spectrum, which leads us to report them as equal.

\subsection{Full dynamics in CFC}

\begin{figure}[h]
    \centering
    \includegraphics[width=\linewidth]{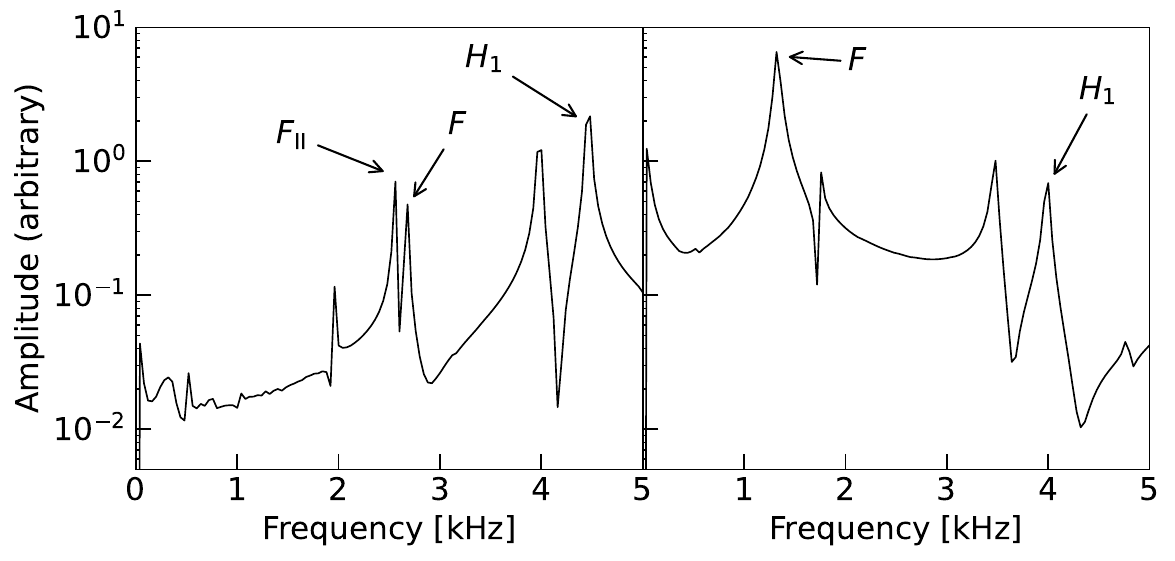}
    \caption{Fourier spectrum of the $l=m=0$ $R_S$ coefficient for the non-axisymmetric B4 simulation in the Cowling approximation (left) and a dynamical spacetime (right).}
    \label{fig:B4_00_spectra}
\end{figure}

Having checked that \texttt{ROXAS} produces consistent results with other articles within the Cowling approximation, we now relax this assumption and work with the Einstein-CFC system solved dynamically (see \cite{Servignat:2024loh} for details), using the full potential of the code. In this case, the waveform spectra show a single fundamental mode, unlike the double peak seen in the Cowling approximation. We can see this more clearly by comparing the spectra of the $l=m=0$ coefficient of $R_S$, from which the fundamental frequencies and their overtones are extracted, as can be seen in Fig.~\ref{fig:B4_00_spectra}. In this figure, we can see the two fundamental-mode peaks around 2.6 kHz in the Cowling simulation, while there is a single fundamental-mode peak around 1.3 kHz in the CFC simulation. This confirms that the secondary fundamental mode is an artifact of the Cowling approximation in differential rotation, as was suspected in~\cite{Stergioulas:2003ep}. While previous works as~\cite{Dimmelmeier:2005zk} studied axisymmetric modes without the Cowling approximation, it is the first time this difference is explicitly addressed by comparing both types of simulations.

\begin{figure}[h]
    \centering
    \includegraphics[width=\linewidth]{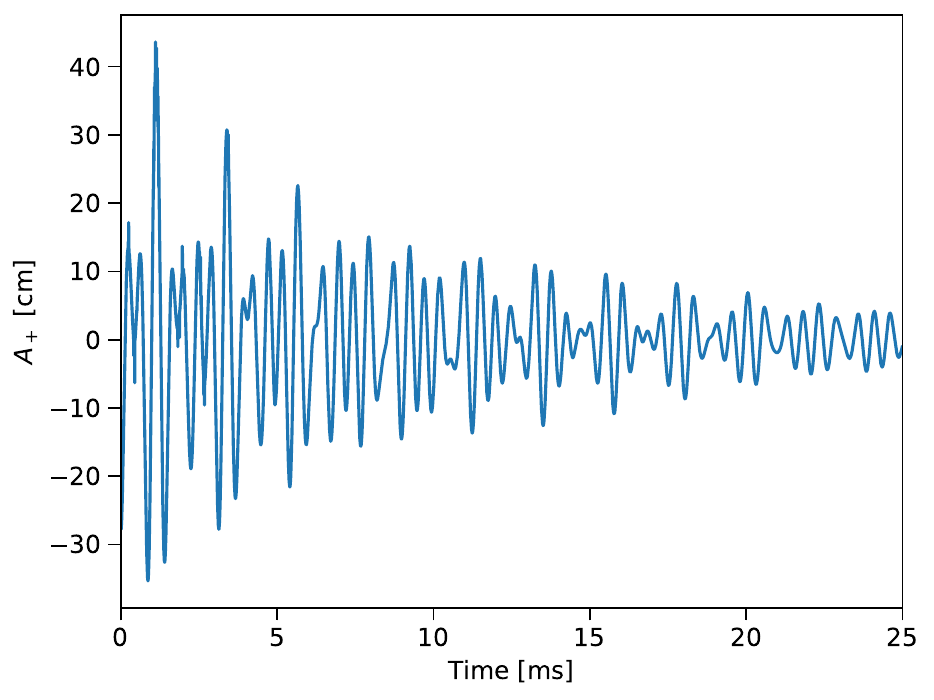}
    \caption{GW strain extracted from the non-axisymmetric B4 simulation in CFC.}
    \label{fig:B4_GW}
\end{figure}

\begin{figure}[h]
    \centering
    \includegraphics[width=\linewidth]{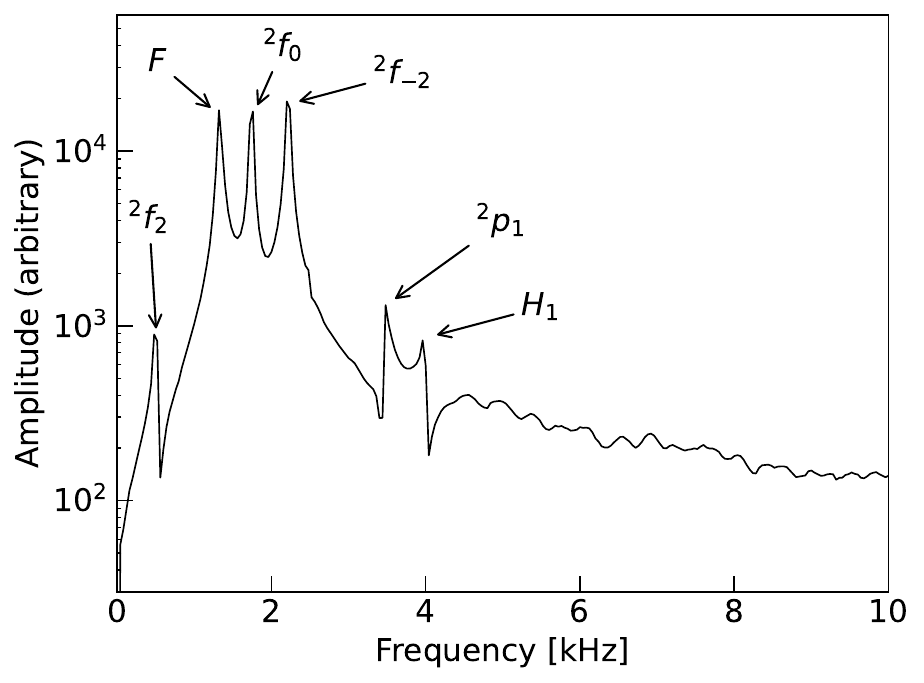}
    \caption{GW spectrum of the non-axisymmetric B4 simulation in CFC.}
    \label{fig:B4_spectrum}
\end{figure}

We show an example waveform in Fig.~\ref{fig:B4_GW}, with its spectrum in Fig.~\ref{fig:B4_spectrum}, corresponding to the non-axisymmetric B4 simulation in CFC. In Table~\ref{tab:freqs}, we report the frequencies for the same axisymmetric and non-axisymmetric modes as those described in the previous section, this time for the B sequence in a dynamical spacetime. The only exception is the secondary fundamental mode, which is absent in this case. Once again, the axisymmetric modes in Table~\ref{tab:freqs} correspond to the axisymmetric simulations, while we use the non-axisymmetric ones to check consistency for these modes and to extract the non-axisymmetric frequencies. The maximum relative error between the axisymmetric modes of both types of simulations is of 6.5\%, which corresponds to $H_1$ for B9. This is an outlier caused by a higher-frequency peak ($\approx 4.45$ kHz) which merges with the one corresponding to $H_1$ in the non-axisymmetric simulation. Both of them can be resolved in the axisymmetric one, as would likely be the case with a longer non-axisymmetric simulation. Other than that, the largest relative error is 2.4\%.

\begin{table*}[h]
    \centering
    \begin{tabular}{|c||c|c|c|c|c|c||c|c|c|c||c|c|}
        \hline
        Model & $F$ & $H_1$ & ${}^2f_0$ & $^2p_1$ & $^2f_2$ & $^2f_{-2}$ & $\delta F$ & $\delta H_1$ & $\delta\,{}^2f_0$ & $\delta\,{}^2p_1$ & $t_{\rm ax}$ & $t_{\rm 3D}$ \\\hline
        B0 & 1.443 & 3.953 & 1.595 & 3.714 & 1.564 & 1.564 & 1.06 & 0.46 & 0.58 & 0.31 & 25.0 & 25.0 \\
        B1 & 1.429 & 3.915 & 1.634 & 3.701 & 1.118 & 1.945 & 1.57 & 0.32 & 0.34 & 0.32 & 25.0 & 25.0 \\
        B2 & 1.392 & 3.914 & 1.674 & 3.647 & 0.881 & 2.072 & 1.37 & 0.33 & 0.27 & 0.52 & 25.0 & 25.0 \\
        B3 & 1.359 & 3.953 & 1.718 & 3.572 & 0.681 & 2.159 & 1.99 & 0.27 & 0.51 & 0.33 & 25.0 & 25.0 \\
        B4 & 1.327 & 4.003 & 1.759 & 3.482 & 0.514 & 2.216 & 3.12 & 0.27 & 0.70 & 0.22 & 25.0 & 25.0 \\
        B5 & 1.287 & 4.073 & 1.800 & 3.395 & 0.320 & 2.276 & 4.03 & 0.02 & 0.63 & 0.14 & 25.0 & 25.0 \\
        B6 & 1.246 & 4.122 & 1.840 & 3.292 & 0.138 & 2.312 & 5.74 & 0.10 & 1.18 & 0.35 & 25.0 & 24.4 \\
        B7 & 1.210 & 4.193 & 1.881 & 3.199 & $-0.055$ & 2.353 & 12.36 & 0.33 & 1.48 & 0.36 & 25.0 & 19.3 \\
        B8 & 1.172 & 4.224 & 1.921 & 2.990 & $-0.235$ & 2.392 & 8.54 & 0.28 & 1.10 & 3.65 & 24.1 & 17.1 \\
        B9 & 1.130 & 4.177 & 1.960 & 2.912 & $-0.471$ & 2.459 & 19.60 & 6.54 & 2.23 & 3.83 & 20.4 & 14.9 \\\hline
    \end{tabular}
    \caption{Frequencies (kHz) extracted from the B sequence simulations and relative errors (\%) of the axisymmetric modes with respect to the results in~\cite{Dimmelmeier:2005zk}. The last two columns indicate the total simulation times (ms) of the axisymmetric and non-axisymmetric simulations.}
    \label{tab:freqs}
\end{table*}

\begin{figure}[h]
    \centering
    \includegraphics[width=\linewidth]{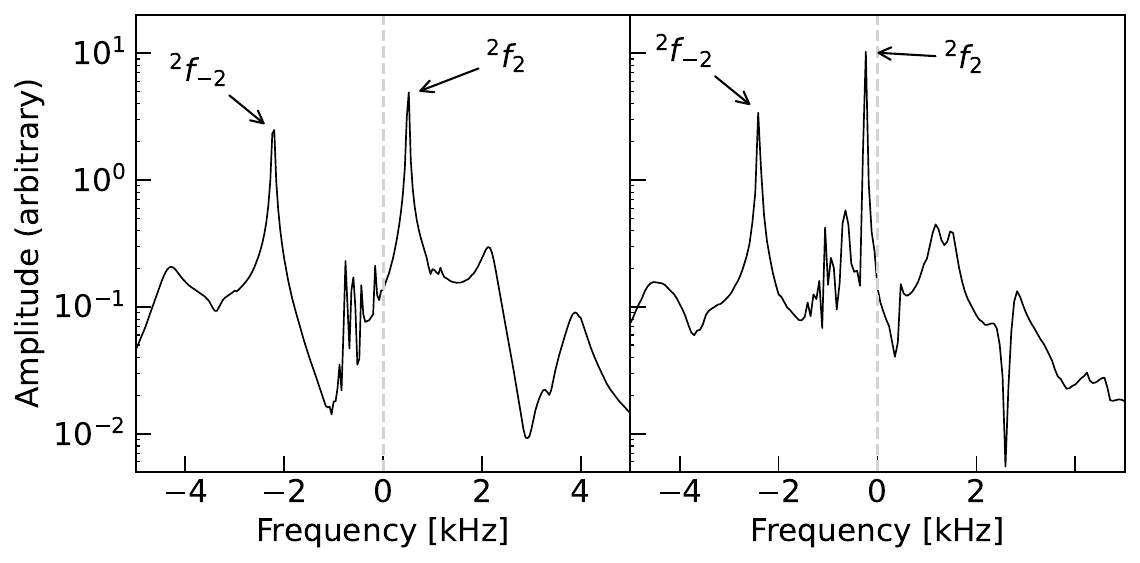}
    \caption{Fourier spectrum of the complex $l=m=2$ $R_S$ coefficient for the non-axisymmetric B4 (left) and B8 (right) simulations in CFC.}
    \label{fig:22_spectra}
\end{figure}

The non-axisymmetric modes are reported here for the first time to our knowledge. The $^2f_2$ frequency becomes negative from B7 onward, indicating that this mode rotates in the opposite direction with respect to the previous cases, in the star's co-rotating reference frame. For low rotation frequencies, the $(2,2)$ mode corresponds to a prograde rotation, while the $(2,-2)$ is retrograde. The regime where the $(2,2)$ mode is also retrograde is called CFS (Chandrasekhar-Friedman-Schutz) instability~\citep{Chandrasekhar:1970,Friedman:1978,Friedman:1978hf}, as reported in~\cite{Kruger:2009nw} within the Cowling approximation and in~\cite{Zink:2010bq,Kruger:2020ykw} for rigid rotation. With our approach, both rotation directions can be distinguished with a Fourier transform of the complex $l=m=2$ $R_S$ coefficient. We can see the different behaviors in Fig.~\ref{fig:22_spectra}.

On the other hand, the axisymmetric modes can be compared to the values quoted in~\cite{Dimmelmeier:2005zk}, in which axisymmetric, dynamical simulations were performed in CFC. The relative errors between this work and ours are also included in Table~\ref{tab:freqs}.

\begin{figure}[h]
    \centering
    \includegraphics[width=\linewidth]{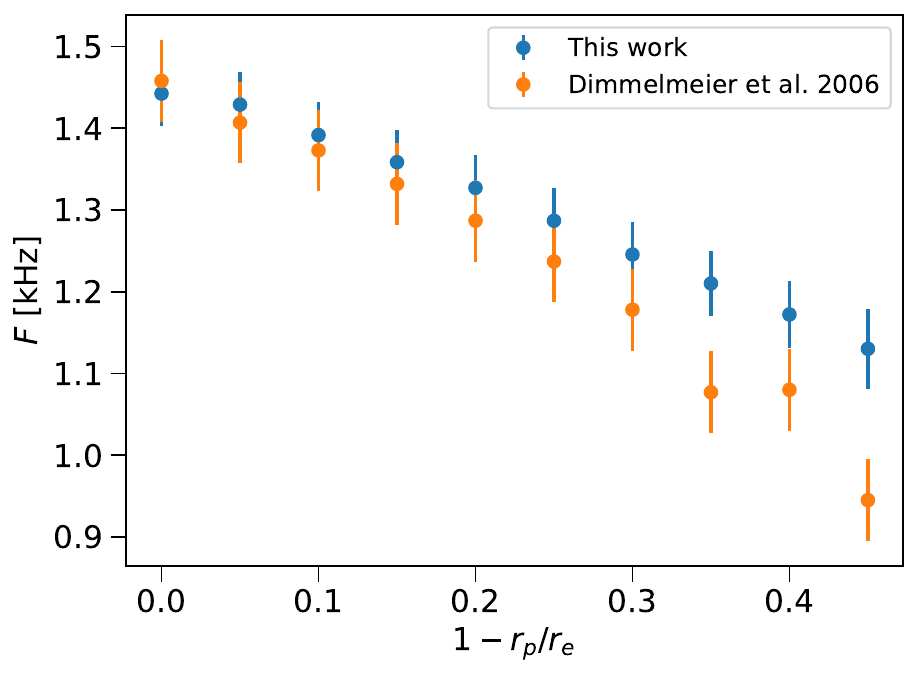}
    \caption{Fundamental frequencies obtained for the B sequence simulations and comparison with those in~\cite{Dimmelmeier:2005zk}. The error bars are the inverse of the simulation times, which correspond to the frequency resolutions.}
    \label{fig:f_modes}
\end{figure}

We can see that the agreement of $H_1$, $^2f_0$ and $^2p_1$ is of order 1\% or lower in most of the cases. For the $H_1$ corresponding to B9, a similar issue to ours might have also arisen in~\cite{Dimmelmeier:2005zk}, as their reported frequency is closer to the misleading 4.45 kHz found in our simulations, leading to a similar relative error of $6.5\%$. The fundamental frequency $F$, however, shows larger discrepancies, especially in the fastest rotating cases. Taking into account the uncertainties purely associated to the frequency resolution limited by the total simulation time (see Fig.~\ref{fig:f_modes}), there is only a clear tension for the B7 and B9 simulations. However, the fundamental frequencies in~\cite{Dimmelmeier:2005zk} show a faster decaying trend than ours. The origin of these discrepancies is unclear at this point, as the codes, resolutions and simulation times of both works are different, which requires a more detailed investigation. Note that the cause is not the CFC approximation, which is also used in~\cite{Dimmelmeier:2005zk}. While a comparison with frequencies obtained directly from GR is not possible here, it was done for the rigidly rotating case in~\cite{Servignat:2024loh} with excellent agreement, showing the reliability of CFC for NS oscillation-mode studies.

\section{Conclusions}
\label{sec:conclusions}

In this work, we have presented an update of the publicly available code \texttt{ROXAS}, which can now simulate differentially rotating stars, together with several additional improvements. We have run a series of both non-axisymmetric and axisymmetric simulations, first in the Cowling approximation and then with a dynamical spacetime in CFC. We have verified their consistency with different references and obtained excellent overall agreement for both axisymmetric and non-axisymmetric mode frequencies. In addition, we have shown that, as conjectured in \cite{Stergioulas:2003ep}, the secondary fundamental peak appearing in simulations within the Cowling approximation does not appear in dynamical spacetimes under the same initial conditions.

Furthermore, the non-axisymmetric frequencies $^2f_2$ and $^2f_{-2}$ in CFC are reported here for the first time, illustrating the ability of \texttt{ROXAS} to explore a wider class of configurations and generate novel results. We also emphasize that these simulations are performed with a lightweight code that runs efficiently on standard computers within short computation times. For radial (1-dimensional) simulations, \texttt{ROXAS} typically performs about 5 times faster~\citep{Servignat:2022ovx} than the reference core-collapse and NS oscillation code \texttt{CoCoNuT}~\citep{Dimmelmeier:2004me}, an effect that should be enhanced for axisymmetric (giving roughly a factor $5^2 = 25$) and non-axisymmetric (factor $5^3 = 125$) simulations.

Together with~\cite{Servignat:2024loh}, this work represents an initial step towards a versatile tool for studying perturbed, rotating NSs with moderate computational resources. Building on this, future works will include differential rotation profiles which are more motivated from binary NS simulations, such as those in~\cite{Uryu:2017obi}, as well as more realistic equations of state, which will require further extensions of the present framework.

\section*{Acknowledgments}

The authors acknowledge support from the Agence Nationale de la Recherche (ANR) under contract ANR-22-CE31-0001-01.

\bibliographystyle{aa} 
\bibliography{main}

\end{document}